\documentclass[twocolumn,showpacs,preprintnumbers,amsmath,amssymb,aps]{revtex4}

\usepackage{graphicx}
\usepackage{dcolumn}
\usepackage{bm}

\def\N{{\cal N}}

\def\S_1{{\widetilde {S_1}}}

\def\R{{\mathbb R}}

\def\tr{{\rm tr}}

\def\Z{{\mathbb Z}}

\def\Dslash{{\rlap{\raise 1pt \hbox{$\>/$}}D}}

\begin{document}

\preprint{SLAC-PUB-12726}

\title{ Abelian duality, confinement, and chiral symmetry breaking in 
 QCD(adj)  
\\ }
\author{Mithat \"Unsal}
\affiliation{%
SLAC  and Department of Physics, Stanford University, CA, 94025 
}%

\date{\today}

\begin{abstract}
We analyze the vacuum structure of  $SU(2)$  QCD with  multiple massless adjoint representation fermions   formulated on  a small spatial $S^1 \times \R^3$.  The absence of thermal 
fluctuations,  and the fact that quantum fluctuations favoring the vacuum with unbroken center symmetry 
in a weakly coupled  regime  renders   the interesting dynamics of these theories analytically calculable. 
Confinement, the area law behavior for large Wilson loops,  and  the generation of the mass gap in the gluonic sector are  shown analytically. 
By abelian duality transformation, the long distance effective theory of QCD is mapped into an amalgamation of $d=3$ dimensional Sine-Gordon and NJL models.  
The duality necessitates  going  to IR first.  In this regime, theory exhibits confinement 
without continuous chiral symmetry breaking. 
However, a flavor singlet chiral condensate (which breaks a discrete chiral symmetry) 
persists at arbitrarily 
small $S^1$. Under the reasonable assumption   that  the theory on $\R^4$ exhibits chiral symmetry 
breaking,  there must  exist a zero temperature 
chiral phase transition in the absence of any change in spatial
center symmetry realizations.
\end{abstract}

\maketitle

This letter aims to address confinement, mass gap in gluonic sector, and chiral symmetry realizations in certain locally four dimensional, asymptotically free 
 QCD-like gauge theories. To date,  confinement is understood quantitatively in a subclass of non-abelian gauge theories, such as Polyakov's model on $\R^3$ 
\cite{Polyakov:1976fu},
 and in mass deformation of   $\N=2$ SYM on $\R^4$ \cite{Seiberg:1994rs}.   
 The  common feature of both theories  is the presence of elementary scalars in the
  defining Lagrangian and gauge symmetry breaking. In both 
 cases, confinement occurs via monopole condensation.  The QCD-like theories in four dimensions 
 lack elementary scalars, and therefore are conceptually harder. (See 
 Refs.\cite{Greensite:2003bk, Shifman:2007ce, 'tHooft:1999au} for reviews.)
  We hope to provide 
 useful insights into such theories on  $\R^3 \times S^1$,  ($\R^{2,1} 
 \times S^1$  in the Minkowski setting).  
 
In  QCD-like theories formulated on small $S^1 \times \R^3 $ (hence weak coupling), the Wilson line  along the compact direction may be viewed as an adjoint Higgs field. In cases where $S^1$ is thermal,  
the thermal fluctuation causes the center symmetry to break and the theory is in  deconfined phase 
\cite{Gross:1980br}.
It would be nice if the center symmetry was not  broken at small $S^1$.  However, at high temperature, 
 this is not debatable, since thermal fluctuations will necessarily overwhelm  the  center symmetry. 
 We want the benefit of weak coupling (small  $S^1$), and the lack of thermal fluctuations.   Therefore,  
 we consider  zero temperature QCD at   small spatial $S^1 \times \R^3$.  
 Since we do not a priori know what the effect of  zero temperature 
 quantum fluctuations will be, we should be ready for surprises, and novel phenomena.  
 
Remarkably,   QCD with $n_f$ adjoint Weyl fermions [abbreviated as QCD(adj)] with periodic spin connection along the $S^1$    respects its  center symmetry in a weakly coupled regime
 \cite{Kovtun:2007py}. The benefit of  weak coupling is that  gauge symmetry is broken, and unlike   the thermal case, the long distance theory abelianizes. 
By quantizing $SU(2)$  QCD(adj)  on small $S^1 \times \R^3$, we exhibit 
\begin{itemize}
\item[i)]Permanent confinement, the area law behavior for  Wilson loops, 
\item[ii)] Absence of  continuous chiral symmetry breaking,  
\item[iii)] Presence of a  flavor singlet chiral condensate 
which only  breaks the discrete chiral  symmetry, 
\item[iv)] The existence of a mass gap in the gluonic sector, and massless fermions in the spectrum
\end{itemize}
With the  assumption that the theory on large $S^1 \times \R^3$ exhibits chiral symmetry 
breaking,  ii) implies 
\begin{itemize} 
\item[v)] The existence of a chiral phase transition in the absence of any change in center symmetry
\end{itemize}
This is a zero temperature phase transition triggered  solely 
by quantum fluctuations. 

{\bf Perturbation theory and spatial center symmetry:}  
The action of  $SU(2)$ QCD(adj) defined on $\R^3 \times S^1$   is 
\begin{equation}
S= \int_{\R^3 \times S^1} \frac{1}{g^2} \tr \left[ \frac{1}{4} F_{MN}^2 + i \bar \lambda^I \bar 
\sigma^M D_{M} \lambda_I \right]
\label{eq:Lagrangian}
\end{equation}
where $\lambda_I= \lambda_{I,a} t_a,   a=1, 2,3$ is Weyl fermion in adjoint representation, 
$F_{MN}$ is the nonabelian gauge field strength, and  $I$ is the flavor index.  
Classically, the theory
 possess an  $U(n_f)$ flavor symmetry  whose $U(1)_A$ part is anomalous. 
The symmetry of the  quantum theory  is $SU(n_f) \times \Z_{4n_f}$. The quantum 
theory has the  dynamical strong scale $\Lambda$, which arises via dimensional transmutation, 
and  is given  by $\Lambda^{b_0} = \mu^{b_0} e^{-4 \pi^2/g^2(\mu)}$ where $\mu$ is the 
renormalization 
group scale and $b_0= (11-2n_f)/3$.    We consider $1<n_f \leq 4 $ so that asymptotic freedom is preserved and the theories are nonsupersymmetric. 

At small $S^1$ $(L \Lambda \ll1)$,  due to asymptotic freedom,  
the gauge coupling is  small and a  perturbative Coleman-Weinberg analysis  is 
reliable \cite{Coleman:1973jx}.  
The minimum of the gauge field action corresponds to the vanishing 
field strength, and constant but arbitrary values of the Wilson line  
\begin{equation} 
U= \left( \begin{array}{cc}
       e^{i \phi} & 0 \\
         0 & e^{-i \phi}
          \end{array} \right) \; .
\end{equation}
Integrating out the heavy KK-modes  along the $S^1$ circle, $|\omega_n|\geq \omega_1$ where
$\omega_n= \frac{2 \pi}{L}n, n \in \Z$, induce a nontrivial effective potential for $U$, given by (up to an
uninteresting constant)
\begin{equation}
V_{\rm eff}[U]= - \frac{(n_f -1)}{ 24 \pi^2 L^4} [2 \phi]^2 ( [2 \phi]- 2 \pi)^2
\label{Eq:CW}
\end{equation}
where $\phi \equiv \phi + 2 \pi$ is a periodic variable. The potential is bounded. 
The action for the classical zero modes reduce to
\begin{eqnarray}
S= \int_{\R^3} \frac{L}{g^2} \tr  &\Big[ \frac{1}{4} F_{\mu \nu}^2 + \frac{1}{2} (D_{\mu} \Phi)^2 +  
g^2 V(| \Phi |) \cr
&+i \bar \lambda^I (\bar \sigma^{\mu} D_{\mu} + \bar \sigma_4 [ \Phi, \; ]) \lambda_I \Big]
\label{Eq:compact}
\end{eqnarray}
The minimum of the potential $V_{\rm eff}$
 is located at $|\Phi|\equiv \phi= \frac{\pi}{2}$, 
hence $ U= {\rm Diag}(e^{i \pi/2}, e^{-i \pi/2}) $.
Since $\tr U = 0$, the $\Z_2$
center symmetry is preserved. By Higgs mechanism, the
gauge symmetry is broken down as 
\begin{equation}
SU(2) \rightarrow  U(1) 
\end{equation}
{\it Remark:} This is unlike thermal QCD(adj), in which the
minimum of the potential for the thermal Wilson line is 
located at $U= \pm 1$ and center symmetry is spontaneously
broken. The gauge symmetry remains  unbroken, and the theory
reduces to non-abelian $d=3$ dimensional pure     $SU(2) $  
Yang-Mills  theory     at long distances. 

Due to the gauge symmetry breaking via an ``adjoint Higgs field", the neutral fields aligned with $U$ along  the Cartan subalgebra ($A_{3,\mu}, \lambda^{I}_{3}$)   remain massless, and off-diagonal components acquire mass, given by the separation between the eigenvalues of the Wilson line 
\begin{equation}
m_{W^{\pm}}= m_{\lambda^{I, \pm}} = (\phi_1 - \phi_2)/L = \pi/L
\end{equation}
where $\pm$ refers to the charges under unbroken $U(1)$. Since
higher order perturbative effects cannot change
the conclusion about the center symmetry realizations,
within perturbation theory,  the low energy theory is a $d=3$  dimensional abelian $U(1)$ gauge theory 
with $n_f$  massless fermions with a free action 
 \begin{equation}
S= \int_{\R^3 } \frac{L}{g^2} \left[ \frac{1}{4} F_{3,\mu \nu}^{2} + i \bar \lambda^I_{3} \bar 
\sigma^\mu \partial_{\mu} \lambda_{3,I} 
\right]
\label{eq:pert}
\end{equation}
At distances shorter than $L$, the coupling constant flows
according to the four dimensional renormalization group.
Since the heavy   $W^{ \pm}, {\lambda^{I, \pm}}$ which are charged under 
$U(1)$  decouple from the long distance physics at scale $L$ and above,   the coupling 
  constant ceases to run  at $1/L \gg \Lambda$  much before the  strong coupling sets in.  
In perturbation  theory, this is the whole story.

Nonperturbatively  though, the free infrared fixed point  is unstable.   This follows from the 
existence of monopoles. The distinction between the free $U(1)$ theory, and the theory 
with monopoles  is that in the latter the $U(1)$ symmetry enhances to 
the whole $SU(2)$   at the monopole cores.  
Below, we will demonstrate that, due to nonperturbative effects,   
 QCD(adj) exhibits confinement.  However, we first have to take a detour and answer 
 the following question: 

{\bf Is this  PolyakovÕs model with  adjoint fermions on $\R^3$?}
The action Eq.\ref{Eq:compact} looks  ``almost" like  
the generalization of the 
 Polyakov model  on $\R^3$  in the presence of $n_f$  
Dirac fermions in adjoint representation, with one difference: 
the compact adjoint Higgs field in   Eq.\ref{Eq:compact} 
has to be substituted by a non-compact one.  
\begin{equation} 
V_{\rm eff}^{\rm compact}(|\Phi|) \rightarrow  V_{\rm eff}^{\rm noncompact}(|\Phi|) 
\label{Eq:PolyakovE}
\end{equation}
Such extensions 
are studied in Ref. \cite{Affleck:1982as} and do  {\bf not}  exhibit 
  confinement.  
 What is going on?    What is the conceptual  difference between the two which 
 results in such drastically different physics?  
 
The   simplest explanation is through the symmetries.
 Let me give the microscopic explanation.    
 In odd dimensions,  there is no chiral anomaly. 
 Hence, the theory on $\R^3$ has a genuine $U(n_f)$ flavor 
 symmetry whose $U(1)$ part is just fermion number.   
Since the  QCD(adj)  theory on small $S^1 \times \R^3$  is locally 
four dimensional, it  has only $SU(n_f) \times \Z_{4n_f} $ symmetry, where 
 $\Z_{4n_f}$ is the anomaly free part the anomalous $U(1)_A$ symmetry.
 All the effective long distance theories must obey the symmetries of their microscopic 
 origin. Hence, in particular, the $U(1)$ symmetry  will be a symmetry of the effective 
 theory corresponding to the extension of Polyakov's model, and $\Z_{4n_f} $ will be the one  
 of QCD(adj).  
 
In both theories, nonperturbatively, there exists   topologically stable, semiclassical field configurations, i.e, monopoles. Their existence follows from the gauge symmetry breaking. Since the second homotopy group $\pi_2[SU(2)/U(1)] = \pi_1[U(1)]= \Z$, 
 there is one type of
ÒalmostÓ BPS monopole.  
 In contradistinction,  in QCD(adj),    there is also a KK-monopole which may be interpreted either due to compactness (or equivalently, due to the fact that the underlying theory is defined on $S^1 \times \R^3$.) 
 Neither monopole is exactly BPS, and their action receives small $O(g^2)$ correction 
 $S_i= \frac{4\pi^2}{g^2}(1+ O(g^2))$
 which we will neglect. One important point is that the magnetic charge of the KK-monopole is 
opposite to that of the  BPS monopole. There are also antimonopoles.

Now, let us review the construction of  the long distance effective theory for the extension of Polyakov's model, see   Ref. \cite{Affleck:1982as} for a full discussion. Let 
$\sigma$ denote the scalar  dual to the photon obtained via the abelian duality.  (The infrared physics 
is easier to 
describe   in the dual description.)  In the background of the   BPS monopole,     due to Callias index theorem  there exists  $2n_f$ fermionic zero modes    \cite{Callias:1977kg,Jackiw:1975fn}.
 Hence, a manifestly $SU(n_f)$  invariant monopole induced 
 fermion vertex should involve  $\det \psi^I \psi^J $.  This vertex is noninvariant under the $U(1)$
 fermion number,   and this  can be cured  by coupling to the  dual photon. 
Generalizing the result of Ref.\cite{Affleck:1982as} to multiflavor $(n_f >1)$,   we obtain  the long distance  effective Lagrangian as 
\begin{equation} 
L_{\rm eff}^{\rm n.c.} = \frac{1}{2} (\partial \sigma)^2 + i \bar \psi^I \gamma^{\mu} \partial_{\mu} \psi_I + 
a e^{-S_0}( e^{i  \sigma} \det \psi^I \psi^J + {\rm c.c.})
\end{equation} 
This is respectful to all the symmetries of the underlying
theory: Most importantly, the fermion number symmetry
\begin{equation}
\psi^I \rightarrow e^{i  \alpha} \psi^I, \;\; \bar \psi^I \rightarrow e^{-i  \alpha} \bar \psi^I, \;\; 
 \sigma \rightarrow \sigma -  2n_f \alpha \; .
\end{equation}
The    $U(1)$ symmetry prohibits any kind of
mass term (or potential) for the dual photon.  As shown in   Ref.\cite{Affleck:1982as},  the
$U(1)$  symmetry is spontaneously broken down to $\Z_{2n_f}$  due to the monopole induced 
$\det\psi^I \psi^J$ term. (This can be seen by expanding the photon around $\sigma=0$, for example.) 
Hence, there must be a Golstone boson associated  with it. In three dimensions, there is no distinction between scalars and vectors, and the photon is the Goldstone boson of the spontaneously broken fermion number  symmetry. Since the photon does not acquire a mass, it remains
infinite range, i.e., there is no confinement.

Notice that in the effective lagrangians, we will always use  dimensionless coordinates, fields ($\sigma$ and $\psi$)  all measured in units of $L$. We will also not calculate various one loop factors in these lagrangians, such as $a,b, c$ which are calculable, but inessential for our conclusions.  

{\bf Abelian duality and dual QCD:}
In  QCD(adj), the $U(1)_A$ which may potentially prevent the photon from acquiring mass is not a  real  symmetry. Its anomaly free incarnation is $\Z_{4n_f}$ which can not prevent a mass term for
the dual photon. Let us see this in detail.
Just like the BPS monopole, 
the KK-monopole will also induce a determinantal fermion vertex, 
 $e^{-i  \sigma} \det \psi^I \psi^J$ where the relative minus sign reflects the fact that KK and BPS monopoles   carry opposite  charges. 
The 
combined effect of BPS and KK monopoles is   $\cos ( \sigma)  \det \psi^I \psi^J$.
This vertex is manifestly invariant under $SU(n_f)$, and  respects the discrete symmetry 
\begin{equation}
\psi^I \rightarrow e^{i  2\pi/(4n_f)} \psi^I, \;\;  \qquad 
 \sigma \rightarrow \sigma + \pi 
\end{equation}
The presence of KK monopoles makes it impossible for the interaction vertex to be invariant under a continuous $U(1)$. 
This is how the  $d=3$ dimensional action ``sees"  that it has a hidden forth dimension, and  it is genuinely different from a locally  three dimensional theory.  The simplest potential term for $\sigma$ allowed by the $\Z_2$ shift  symmetry is $[e^{-S_0}\cos \sigma]^2 \sim e^{-2S_0} \cos 2 \sigma  $. Hence, 
the long distance effective 
theory which describes the dynamics of QCD(adj) on small $S^1 \times \R^3$ is 
\begin{eqnarray}
&&L^{\rm dQCD} = \frac{1}{2} (\partial \sigma)^2 -  b\;  e^{-2S_0}\cos 2 \sigma  \cr 
+ && i \bar \psi^I \gamma_{\mu} \partial_{\mu} \psi_I   
 + c \; e^{-S_0}  \cos \sigma   
( \det_{I, J} \psi^I \psi^J +  \rm c.c.) \qquad 
\end{eqnarray}
which is manifestly invariant under $SU(n_f) \times \Z_{4n_f}$  symmetry of the original theory. 

{\bf Mass gap in the gauge sector:}  The small fluctuations 
around one of the  minima of the $\cos 2\sigma$  potential 
shows  that the dual photon acquires a mass, proportional to $e^{-S_0}$.  
This is  the Debye
mass in the classical plasma.  In terms of the gauge theory, 
it is the inverse characteristic size of the chromoelectric 
flux tube. For
$n_f$  flavor theory, it  is given by
\begin{equation}
m_D \sim \Lambda(\Lambda L)^{b_0 -1} =  \Lambda(\Lambda L)^{(8-2n_f)/3} \; .
\end{equation}
This is a remarkable result. It exhibits that the gauge sector of the QCD(adj) theory is quantum mechanically gapped due to non-perturbative effects.  Since the chromoelectric fields become
 short range, this also implies confinement.  
%

{\bf Area law of confinement and monodromy:} We wish to exhibit the area law of confinement by calculating  Wilson
loops in the half-spin representations. 
  The
representation of the Wilson loops   under the center group $\Z_2$ 
are in one to one correspondence with the  monodromies, 
$\int_C d \sigma$ in the dual theory  \cite{Deligne:1999qp}.   (Both are representation of 
$\Z_2$.) 
Evaluating the Wilson loops in a representation with odd or even  
 $\Z_2$ center group charge  assignment 
   in the original theory translates into finding the field configurations for the dual scalar with 
monodromies  equal to $ \pi $ or $0$  (mod $2\pi$), respectively.  
Therefore, we need to classify the vacuum
states in the dual theory, and the soliton configurations
interpolating between them.
The Sine-Gordon potential has two
gauge inequivalent vacua 
   \begin{equation}
  |\Omega_0 \rangle \equiv    |\Omega_{0+ 2 \pi k} \rangle, \qquad  
  |\Omega_1 \rangle \equiv    |\Omega_{\pi  + 2 \pi k } \rangle, \;\;  k \in \Z
    \end{equation}
  Therefore, the expectation values of the Wilson loop falls into two categories, for 
  half-integer and integer spin representation.
Let $H$ denote the Hamiltonian of the dual theory: The Hilbert space interpretation 
of   Polyakov's result is 
  \begin{eqnarray}
\lim_{A(\Sigma) \rightarrow \infty}  \langle W_{\rm odd(even)}(C) \rangle |_{C= \partial \Sigma} =   
\langle \Omega_{1(0)} | e^{-z H}| \Omega_0 \rangle 
\end{eqnarray}
where $z$ is Euclidean time, and interpolates between $[-\infty, \infty]$. The expectation value of arbitrarily large Wilson loops   (where $\Sigma$ is  $\R^2$ filling)
  are equal to tunneling amplitudes in the dual theory.  Formally,  the  tunneling amplitude on  $\R^2 \times \R$ is $e^{-{\rm Area}(\R^2) S^{*}}$ where $S^{*}$ is the least action associated with $x, y  \in \R^2$ independent soliton (kink) solution.  The kink is 
localized within the $m_D^{-1}$ proximity of the surface $\Sigma$. This translates into 
 the magnetic  charge carriers   forming a dipole layer in the vicinity of the surface $\Sigma$ to   prevent the penetration of the external magnetic field into the magnetic conductor, which is the vacuum of QCD(adj) from   Euclidean viewpoint.  Since 
 $
\langle W_{\rm odd}(C) \rangle =  e^{-T A(\Sigma)}$ where $T$ is string tension, $T\equiv S^{*}$,  
\begin{eqnarray}
\qquad T \sim  \Lambda^2(\Lambda L)^{b_0 -2} =  \Lambda^2(\Lambda L)^{(5-2n_f)/3} \; .
\end{eqnarray}
This exhibits the area law of permanent quark 
confinement  in QCD(adj) in the $L \Lambda \ll 1$ regime. 
We expect the tension  to saturate to a c-number times $\Lambda^2$ for $L\Lambda > 1$.

On the other hand, $0\equiv 2 \pi$  monodromy can be induced
by no-soliton, and even soliton  
sector of the dual theory. Hence,   $\langle W_{\rm even}(C) \rangle = 1 + 
O(e^{- 2 T{\rm Area}(\Sigma)})$, and no area law as expected. (In the strongly coupled regime, 
this must become perimeter law.)

{\bf Chiral symmetry realizations:}
At small $S^1$,  
we will argue that
the only broken symmetry is the discrete chiral symmetry (which is intertwined with 
$\Z_2$ shift symmetry of photon and we already showed this),   and no continuous 
chiral symmetry is broken. 
Consider for simplicity the theory with $n_f=2$ flavors.  Since $\sigma$ is massive, at low energies, 
the appropriate lagrangian (around $\sigma=0$ ) is  
 \begin{equation}
  L_{\rm NJL}=  
 i\bar \psi^I \gamma_{\mu} \partial_{\mu} \psi^I   + c  e^{-S_0}  
 ( \det_{I, J} \psi^I \psi^J +  \rm c.c.) 
\label{Eq:NJL}
\end{equation}
NJL type \cite{Nambu:1961tp}. 
The action is invariant under the flavor symmetry $(SU(2) \times \Z_4)$.   We wish to know 
whether it is spontaneously broken. 
The $d=3$ dimensional NJL models  has generically   two
phases depending on the coefficient of the fermion self-interaction $g$ in units of cut-off.
In the $g \sim 1$  regime, NJL
models exhibit a chiral transition from a chirally symmetric 
phase at weak coupling to a chirally asymmetric phase
in the strong coupling $g > 1$. (See, the review Ref.\cite{Rosenstein:1990nm}).
Our  dimensionless coupling constant is $g \sim e^{-S_0}$, which is a tiny number.
Hence, the chiral symmetry must  be unbroken, and there must be massless fermions (protected by chiral symmetry) in the spectrum 
 within the region of validity of our long distance effective theory,  ($L \Lambda \ll 1$).   We believe this is true for $n_f>2$, as well. 

The  unbroken continuous chiral symmetry does not exclude
the presence of flavor singlet chiral condensates. Such
an operator is $ \det \tr \lambda^I \lambda^J$.   
The calculation is slightly technical, but we can estimate 
it on physical grounds. It must be proportional to $e^{-S_0} L^{-3n_f}$. 
The $e^{-S_0}$ reflects the fact that it is due to the one monopole sector. And the  deceptive 
UV divergence in the effective lagrangian  Eq.\ref{Eq:NJL} is cut-off  by the finite 
size of the  monopoles in the full theory.  Factoring out 
$\Lambda^{3n_f} $, the expected behavior on $\R^4$, we obtain
\begin{equation}
\langle \Omega_k | \det  \tr  \;  \lambda^I  \lambda^J |\Omega_k \rangle   
\sim  \Lambda^{3n_f} 
 (\Lambda L)^{ \frac{11}{3}(1-n_f)}  e^{\frac{i 2 \pi k}{2}} 
\label{Eq:condensate2}
\end{equation} 
and  the  phase is $\Z_2$ valued.  
Hearteningly, this produces the correct $L$  independence in the $n_f=1$ case, which is just 
$\N=1$ SYM \cite{Davies:2000nw}, and two isolated vacua.  
   
 At large $S^1$ (and $\R^4$), the common lore is that 
the chiral symmetry is spontaneously broken down to $SO(n_f) \times \Z_2$, hence there are two isolated coset spaces each of which is $SU(n_f)/SO(n_f)$,   and distinguished from each other  by the  
phase of $\langle \det  \tr  \;  \lambda^I  \lambda^J \rangle  \in \Z_2$.
This implies QCD(adj) must possess a (nonthermal) chiral phase transition in the absence of any change in its spatial center symmetry realization.  (The existence of the transition can be proven 
by  considering the theory on $T^2 \times \R^2$, and using Coleman's theorem \cite{Coleman:1973ci}
in small $T^2$ limit.) 
This is a quantum phase transition at absolute 
zero temperature, of  which there are    many examples  in condensed matter physics. 

{\it Remark:} We believe the naive extrapolation of the NJL Lagrangian 
Eq.\ref{Eq:NJL} will also exhibit this transition, and moreover, the transition will take place in an 
expected regime of QCD.   However, this will happen outside the region of validity of our effective theory.  
Consequently, this does not tell us that monopole induced vertex is the dynamical origin of continuous
chiral symmetry breaking. 
   
{ \bf Conclusions: } 
The absence of thermal fluctuations 
and the fact that quantum fluctuation favoring the  unbroken center symmetric vacuum in the weakly coupled regime   is the key which makes   nonperturbative dynamics   
of   QCD(adj)  formulated on spatial $S^1\times  \R^3 $
 analytically tractable.   This provides us one of the few examples of four dimensional gauge theory dynamics which can be understood at a  quantitative level, and  we are optimistic of further progress. 
A detailed discussion of microscopic derivations and  $SU(N)$ generalization is in preparation.

 I  am grateful to O. Aharony, J. Greensite, J. Harvey, D. B. Kaplan, M. Mulligan,  E. Poppitz, M. Shifman, 
 L. Yaffe  for discussions. This
work was supported by the U.S. Department of Energy Grants DE-AC02-76SF00515.

\bibliography{QCD}

\end{document}